# Einstein the Stubborn: Correspondence between Einstein and Levi-Civita

Galina Weinstein[*]

January 31, 2012

Before developing his 1915 General Theory of Relativity, Einstein held the "Entwurf" theory. Tullio Levi-Civita from Padua, one of the founders of tensor calculus, objected to a major problematic element in this theory, which reflected its global problem: its field equations were restricted to an adapted coordinate system. Einstein proved that his gravitational tensor was a covariant tensor for adapted coordinate systems. In an exchange of letters and postcards that began in March 1915 and ended in May 1915, Levi-Civita presented his objections to Einstein's above proof. Einstein tried to find ways to save his proof, and found it hard to give it up. Finally Levi-Civita convinced Einstein about a fault in his arguments. However, only in spring 1916, long after Einstein had abandoned the 1914 theory, did he finally understand the main problem with his 1914 gravitational tensor. In autumn 1915 the Göttingen brilliant mathematician David Hilbert found the central flaw in Einstein's 1914 derivation. On March 30, 1916, Einstein sent to Hilbert a letter admitting, "The error you found in my paper of 1914 has now become completely clear to me".

## Introduction

In his paper from April 1911, Alberto A. Martínez writes, "Unfortunately, the quality and quantity of editorial notes have been inconsistent among the volumes [of the *Collected Papers of Albert Einstein* (*CPAE*)]. Volume 2 shines with approximates 149 full pages of editorial notes and footnotes (out of 693 total pages). By contrast, Volume 6 has merely 51 such pages (out of 656). Volume 6 deals with Einstein's works from 1914 to 1917".[1] Volume 8 of the *CPAE* suffers from quite the same defect. This Volume deals with Einstein's correspondence from 1914 to 1917.[2] It contains Einstein's letters to Tullio Levi-Civita from Padua, one of the founders of tensor calculus, and the latter's one surviving reply to Einstein's letters.[3] Only one paper was published in 1989 on the correspondence between the two by Carlo Cattani and Michelangelo De Maria, "The 1915 Epistolary Controversy between Einstein and Tullio Levi-Civita".[4]

---

[*] This paper was written while I was a visitor scholar of the Center for Einstein Studies in Boston University.

In This paper I endeavor to supply some missing critical information and by this fill a little gap. However, Martínez general criticism should be taken into account. The correspondence between Einstein and Levi-Civita deserves careful historical research, which hopefully would lead to adding editorial notes to Volume 8 of the *CPAE*.

## 1. Einstein's "angepaßte Koordinantensysteme"

In his 1914 paper, "The Formal Foundation of the General Theory of Relativity",[5] Einstein presented an improved Hole Argument which restricted the covariance of the gravitational field equations. Under the restrictions of the Hole Argument, he posed a coordinate condition on the field equations, and the coordinate systems satisfying this condition were the adapted coordinate systems (Angepaßte Koordinatensysteme) for the gravitational field.

Einstein presented a theorem and gave its proof that supplies the formal basis for his belief that if the coordinate system is an adapted coordinate system, then the gravitational tensor is a covariant tensor. Einstein therefore demanded that the desired field equations and also the gravitation tensor would be covariant only with respect to adapted coordinate systems.

Einstein thought that he managed to fulfill the principle of equivalence (for adapted coordinate systems): the centrifugal force, which acts under given conditions upon a body, is determined by precisely the same natural constant as the effect of the gravitational field, such that we have no means to distinguish a centrifugal field from a gravitational field. He thus interpreted a rotating system as at rest and the centrifugal field as a gravitational field.

Einstein explained the formal steps that lead to his choice of adapted coordinate systems. He started with the "Hamiltonian" function H of the contravariant fundamental tensor $g^{\mu\nu}$ and its first derivatives $\partial g^{\mu\nu}/\partial x_\sigma$, where the latter are called $g^{\mu\nu}_\sigma$ for short.[6]

He then wrote the following equation for the integral J; the integral is extended over a finite part Σ of the continuum (using Einstein's original notation and numbers of equations):

$$(61) \quad J = \int H \sqrt{-g} \, d\tau.$$

Consider a coordinate system $K_1$. Einstein asked for the change ΔJ of J when we go from the system $K_1$ to another system $K_2$, which is infinitesimally different from $K_1$.[7]

He considered the change Δ due to infinitesimal transformation of a certain quantity at some point of the continuum, and found:[8]

(62) $\Delta \sqrt{-g} d\tau = 0$.

And then he obtained an expression for $\Delta H$ in terms of $\Delta g^{\mu\nu}$ and $\Delta g_\sigma^{\mu\nu}$:

1) He first referred to the transformation law,

(8) $A^{\mu\nu\prime} = \sum_{\alpha\beta} \frac{\partial x'_\mu}{\partial x_\alpha} \frac{\partial x'_\nu}{\partial x_\beta} A^{\alpha\beta}$,

and expressed the $\Delta g^{\mu\nu}$ in terms of $\Delta x_\mu^9$ and obtained,

(63) $\Delta g^{\mu\nu} = \sum_\alpha \left( g^{\mu\alpha} \frac{\partial \Delta x_\nu}{\partial x_\alpha} + g^{\nu\alpha} \frac{\partial \Delta x_\mu}{\partial x_\alpha} \right)$

and obtained an additional expression for $\Delta g_\sigma^{\mu\nu}$.

2) He assumed that H was invariant under linear transformations. Therefore, $\Delta H$ vanished when $\frac{\partial^2 \Delta x_\mu}{\partial x_\alpha \partial x_\sigma} = 0$.[10]

Under these assumptions:

(64) $\frac{1}{2} \Delta H = \sum_{\mu\nu\tau\alpha} g^{\nu\alpha} \frac{\partial H}{\partial g_\sigma^{\mu\nu}} \frac{\partial^2 \Delta x_\mu}{\partial x_\sigma \partial x_\alpha}$.

From this and by means of equation (62) he obtained an expression for $\Delta J$:

$\frac{1}{2} \Delta J = \int d\tau \sum_{\mu\nu\sigma\alpha} \frac{\partial H \sqrt{-g}}{\partial g_\sigma^{\mu\nu}} \frac{\partial^2 \Delta x_\mu}{\partial x_\sigma \partial x_\alpha}$.

And from this by partial integration:

(65) $\frac{1}{2} \Delta J = \int d\tau \sum_\mu (\Delta x_\mu B_\mu) + F$.

Einstein wrote for $B_\mu$,

(65a) $B_\mu = \Sigma_{\alpha\sigma\nu} \frac{\partial^2}{\partial x_\sigma \partial x_\alpha} \left( g^{\nu\alpha} \frac{\partial H \sqrt{-g}}{\partial g_\sigma^{\mu\nu}} \right)$,

and also wrote an additional expression for F.

This enabled Einstein to restrict the coordinate systems.[11] He considered the finite portion Σ and the coordinate system K. He imagined a series of infinitely related coordinate systems K', K" and so on, so that $\Delta x_\mu$ and $\frac{\partial \Delta x_\mu}{\partial \Delta x_\alpha}$ vanish at the boundaries. For every infinitesimal coordinate transformation between neighboring coordinate systems of the total coordinate systems K, K', K",… we obtain: F = 0, and the equation for ΔJ becomes :

(66) $\frac{1}{2}\Delta J = \sum_\mu \int d\tau \, \Delta x_\mu B_\mu$.

Among the above systems K', K" and so on, Einstein chose coordinate systems which are "coordinate systems adapted to the gravitational field" (Gratatationsfeld angepaßte Koordinantensysteme).[12] What characterizes these systems?

The equations (65a) become:

(67) $B_\mu = 0$

and hold for these adapted coordinate systems. Einstein wrote that the above equations hold for the adapted coordinate systems because the $\Delta x_\mu$ can be chosen freely inside Σ. This is then a sufficient condition that the coordinate system is adapted to the gravitational field.[13]

In section §14 of his 1914 review paper, Einstein proved a theorem that supplied the formal basis for the claim that, if the coordinate system was an adapted coordinate system, then the gravitational tensor was indeed a covariant tensor.

*Theorem*: If the gravitational field of $g_{\mu\nu}$ is varied by an infinitely small amount, so that $g_{\mu\nu}$ are replaced by $g^{\mu\nu} + \delta g^{\mu\nu}$, where the $\delta g^{\mu\nu}$ shall vanish at the boundaries of Σ, then H becomes H+δH and the action J becomes J + δJ. Then the equation:

(68) $\Delta\{\delta J\} = 0$

always holds whichever way the $\delta g_{\mu\nu}$ might be chosen, provided the coordinate systems ($K_1$ and $K_2$) are *adapted* coordinate systems with respect to the unvaried gravitational field. This means that under the restriction to adapted coordinate systems, δJ is an invariant.[14]

*Proof*: Einstein imagined the variations $\delta g^{\mu\nu}$ to be composed of two parts:

(69) $\delta g^{\mu\nu} = \delta_1 g^{\mu\nu} + \delta_2 g^{\mu\nu}$

$\delta_1 g^{\mu\nu}$ – are taken in a manner so that the coordinate system $K_1$ is not only adapted to the gravitational field of the $g_{\mu\nu}$, but also to the *varied* gravitational field:

$g^{\mu\nu} + \delta g^{\mu\nu}$.

Therefore,

(67) $\mathbf{B_\mu = 0}$

and

(70) $\mathbf{\delta_1 B_\mu = 0.}$

$\delta_2 g^{\mu\nu}$ – are taken without changing the gravitational field, by variation of the coordinate system, variation in the sub-domain of $\Sigma$ in which the $\delta g^{\mu\nu}$ are not 0, and thus $\boldsymbol{\delta_2 B_\mu \neq 0.}$

The superposition of the two variations is determined by 10 mutually independent functions, and thus – so believed Einstein – will be equivalent to an *arbitrary* variation of the $\delta g^{\mu\nu}$. He reasoned that the proof of his theorem would be completed once equation (68) was proven for the two partial variations.[15]

Einstein thought he proved his theorem and deduced the existence of 10 components, which has tensorial character if we limit ourselves to adapted coordinate systems.[16]

After varying infinitesimally the $g^{\mu\nu}$, Einstein rewrote equation (61) in the following form:

$$[(61a)] \quad dJ = \int d\tau \sum_{\mu\nu\sigma} \left\{ \frac{\partial(H\sqrt{-g})}{\partial g^{\mu\nu}} \delta g^{\mu\nu} + \frac{\partial(H\sqrt{-g})}{\partial g^{\mu\nu}_\sigma} \delta g^{\mu\nu}_\sigma \right\},$$

and since: $\delta g^{\mu\nu}_\sigma = \frac{\partial}{\partial x_\sigma}(\delta g^{\mu\nu})$

after partial integration and considering the vanishing of $\delta g^{\mu\nu}$ at the boundary of $\Sigma$,

$$(71) \quad \delta J = \int d\tau \sum_{\mu\nu} \delta g^{\mu\nu} \left\{ \frac{\partial H \sqrt{-g}}{\partial g^{\mu\nu}} - \sum_\sigma \frac{\partial}{\partial x_\sigma} \left( \frac{\partial H \sqrt{-g}}{\partial g^{\mu\nu}_\sigma} \right) \right\}.$$

Einstein proved that under limitation to adapted coordinate systems $\delta J$ was invariant.[17]

He concluded that since $\delta g^{\mu\nu}$ differs from zero only in infinitely small areas, and since $\sqrt{-g}d\tau$ is a scalar, the integral divided by $\sqrt{-g}$ is also an invariant:[18]

$$(72) \quad \frac{1}{\sqrt{-g}} \sum \delta g^{\mu\nu} \mathfrak{G}_{\mu\nu}$$

The tensor $\mathfrak{G}_{\mu\nu}$ is Einstein's 1914 gravitation tensor, and it is equal to the quantity in the brackets of (71):

$$(73) \quad \mathfrak{G}_{\mu\nu} = \frac{\partial H \sqrt{-g}}{\partial g^{\mu\nu}} - \sum_\sigma \frac{\partial}{\partial x_\sigma} \left( \frac{\partial H \sqrt{-g}}{\partial g^{\mu\nu}_\sigma} \right).$$

We thus rewrite (71) in the following form,

$$(71) \quad \delta J = \int d\tau \sum_{\mu\nu} \delta g^{\mu\nu} \mathfrak{G}_{\mu\nu}.$$

Einstein concluded, [19]

"It follows that,

$$[(74)] \quad \frac{\mathfrak{G}_{\mu\nu}}{\sqrt{-g}}$$

under limitation to adapted coordinate systems, and substitutions between them, is a covariant tensor, and $\mathfrak{G}_{\mu\nu}$ *itself is the corresponding covariant V-tensor [tensor density] and according to (73) a symmetric tensor*".

## 2. Tullio Levi-Civita's Objections to Einstein's Proof and Gravitational tensor

Tullio Levi-Civita carefully read Einstein's 1914 paper, and attacked Einstein's theory. He objected to the major problematic element in Einstein's theory, which reflected its global problem: its field equations were restricted to an adapted coordinate system. The major problematic element was Einstein's above theorem and its proof from section §14: these supplied the formal basis for Einstein's belief that if the coordinate system was an adapted coordinate system, then the gravitational tensor [(74)] as a covariant tensor.

In an exchange of letters and postcards that began on March 1915 and ended in May 1915, Levi-Civita presented his objections to Einstein's proof: Levi-Civita could not accept that [(74)] was a covariant tensor for adapted coordinate systems (within a theory which was limited in its covariance). In his correspondence with Einstein he demonstrated to the latter that neither was δJ invariant, nor was [(74)] a covariant tensor for adapted coordinate systems.

Einstein tried to find ways to save his proof by answering Levi-Civita quandaries and demonstrations, and in most of the letters he repeated the same arguments over and over again. Einstein found it hard to give up his proof. However, Levi-Civita gradually showed to Einstein that he tried the impossible.

Cattani and De Maria claimed that "Levi-Civita's criticisms contributed to stimulating and early growth of Einstein's 'dissatisfaction' with his *Entwurf* theory. […] Levi-Civita did not question directly the limited covariance properties of Einstein's *Entwurf* equations; instead, he shot his mathematical darts against the proof of a theorem crucial to Einstein's variational derivation of the *Entwurf* equations and, in particular, contested the covariance properties of the so-called gravitation tensor. After many fruitless attempts to rebut Levi-Civita's criticism and to find a more convincing proof of that theorem, Einstein was obliged for the first time to admit that both the proof of that theorem and its consequences were not correct".[20]

Unfortunately, all of Levi-Civita's letters but one single letter were lost, but (the known) Einstein's letters to Levi-Civita were saved. And this was typical to Einstein. He did not save letters and used to discard them. Small wonder that all of Levi-Civita's letters were lost, but only the one from March 28, 1915 was saved; because that was the letter which Einstein enclosed back on April 2, 1915 in his letter to Levi-Civita "so that I can refer to it without any inconvenience to you".[21]

Nevertheless, one can easily reconstruct the contents of Levi-Civita's letters from the contents of Einstein's replies to his letters, because Einstein recapitulates the contents of Levi-Civita's letters, and afterwards replies to the latter's objections.

### The first letter from March 5, 1915: the average values $\overline{\delta g^{\mu\nu}}$ of the $\delta g^{\mu\nu}$

The correspondence between Einstein and Levi-Civita very likely started at the beginning of March with a letter from Levi-Civita, which was lost. From Einstein's letter from March 5, 1915 to Levi-Civita we learn that the latter read the former's paper very carefully, after which he presented his objection to Einstein's proof from section §14 of his 1914 review paper. Einstein replied and wrote Levi-Civita, "When I saw that you are directing your attack against the theory's most important proof, which I had won by sweat, I was a little alarmed, especially since I know that you have a much better command of these mathematical matters than I. Yet after thorough considerations, I believe to uphold my proof right".[22] And so at this stage Einstein adhered to his adapted coordinate systems.

Levi-Civita intended to show in the second part of his letter with the example H = $g_{11}$ that what Einstein considered as a [covariant] tensor [(74)] was not correct.[23] By this Levi-Civita also showed Einstein that equation (72) of his 1914 review paper was not an invariant.

Einstein answered Levi-Civita, "It is, however, informed at the bottom of page 1069 [of his 1914 paper], that H must be chosen in such a way that it is invariant under *linear* substitutions. Without this precondition, which is indeed fundamental to what follows, formula (65) is not valid". Einstein told Levi-Civita that since $g_{11}$ is not an invariant for linear substitutions that "means that your counterexample does not constitute a refutation of my stated theorem".[24]

In the first part of his letter, Levi-Civita did not accept that under limitation to adapted coordinate systems δJ was invariant. Einstein in his letter to Levi-Civita said that he did not quite understand Levi-Civita's objection; he did not understand why the conclusion drawn from equation (71) ought not to apply: "in the calculus of variations it is always done in the way in which I have done it".[25]

Einstein said that he knows that

(71) $\int d\tau \sum_{\mu\nu} (\delta g^{\mu\nu} \mathfrak{G}_{\mu\nu})$

is an invariant regardless of how the $\delta g^{\mu\nu}$'s are chosen.

Einstein then chose the $\delta g^{\mu\nu}$'s in the following form. The $\delta g^{\mu\nu}$'s differ from zero in an infinitely small area σ inside Σ: Einstein was reducing the region Σ, in which the terms $\delta g^{\mu\nu}$ differed from zero, to an infinitely small area σ.

The $\mathfrak{G}_{\mu\nu}$'s could be treated as constants in the integration. Einstein also set for the domain Σ:

$$\int d\tau = \tau$$

And:

$$\int \delta g^{\mu\nu} d\tau = \overline{\delta g^{\mu\nu}} \tau.$$

Where, $\overline{\delta g^{\mu\nu}}$ signifies spatial average values of the $\delta g^{\mu\nu}$ terms; when reducing the region Σ to σ, in which the terms $\delta g^{\mu\nu} \neq 0$, **the mean values $\overline{\delta g^{\mu\nu}}$ approached a limit.** Then Einstein wrote that in this case it was possible to set instead of δJ in (71),

(71a) $\tau \sum \overline{\delta g^{\mu\nu}} \mathfrak{G}_{\mu\nu}$

And then under the limitation to adapted coordinate systems δJ is an invariant.

Einstein took into consideration that due to the smallness of the domain σ inside Σ, the $\overline{\delta g^{\mu\nu}}$'s transform at one place within this domain σ like the $\delta g^{\mu\nu}$'s, and thus like the $g^{\mu\nu}$'s.[26]

In his 1914 paper Einstein concluded, that since δg^μν differs from zero only in infinitely small areas, and since $\sqrt{-g}d\tau$ is a scalar, then (72) is also an invariant.[27] With his $\overline{\delta g^{\mu\nu}}$'s, differing from zero only in an infinite little area within Σ, Einstein actually arrived at the same conclusion using his above procedure.

Einstein tried to demonstrate to Levi-Civita his mathematical reasoning behind his proof, but at the same time his confidence was somewhat shaken toward the end of the letter. He was not altogether sure that his proof was truly solid, and so he ended his letter by saying, "I beg you, to inform me of your opinion of the proof upon reconsideration".[28]

**The second letter from March 17, 1915: the contravariant tensor A^μν**

On March 17, 1915 Einstein responded to Levi-Civita's last letter. Einstein wanted to correspond in Italian because of culture association to Italy, because Einstein asked Levi-Civita, "I shall be delighted if next time you write me in Italian. I spent as a young man over half a year in Italy, and at that time had the pleasure of visiting the charming little town of Padua, and I look forward now to be able to use my modest Italian language skills".[29]

There was an earlier letter, which Einstein mentioned and he did not respond to. From Einstein's response we understand that Levi-Civita was not satisfied with Einstein's March 5 reply.[30] "Your attack", said Einstein to Levi-Civita is mainly that, in reducing the region Σ, in which the terms δg^μν differ from zero, in general, the mean values $\overline{\delta g^{\mu\nu}}$ do not approach the limit.[31] Einstein's previous example from March 5 proving that δJ is an invariant was thus problematic.

At this stage Einstein ignored Levi-Civita's criticism, and said "I will not enter into the issue of the existence or nonexistence of this limit, since this question is, in my opinion, of no importance; rather, I shall attempt to prove that the conclusion drawn from formula (71) of the paper is correct". And so Einstein stubbornly adhered to his adapted coordinate systems, and could not hear any criticism, even when a mathematical flaw was found in his derivation. This incidence was to later reoccur in his correspondence with Wilhelm de Sitter on the issue of the cosmological model and Much's principle. Einstein adhered to his adapted coordinate systems and the Hole Argument before November 1915 in much the same way as he would later adhere to his cosmological model and constant.[32]

Einstein then referred again to his paper and to his proof that under limitation to adapted coordinate systems δJ was invariant; and he intended to show by a different strategy and way that the conclusions arrived at from his equation (71) are correct.

He wrote in the letter δJ as **δI** and numbered equation (71) as (1),

(1) $\delta I = \int d\tau \sum_{\mu\nu} \delta g^{\mu\nu} \mathfrak{G}_{\mu\nu}$.

He started with the fundamental tensor $g^{\mu\nu}$ being contravariant, so it transforms according to (8):

$$(2)\, g'^{\rho\sigma} = \sum \frac{\partial x'_\rho}{\partial x_\mu} \frac{\partial x'_\sigma}{\partial x_\nu} g^{\mu\nu}$$

And the variations $\delta g^{\mu\nu}$ (having tensorial character) transform according to the same equation,

$$(2a)\, \delta g'^{\rho\sigma} = \sum \frac{\partial x'_\rho}{\partial x_\mu} \frac{\partial x'_\sigma}{\partial x_\nu} \delta g^{\mu\nu}$$

He multiplied (2a) by $\sqrt{-g'}d\tau' = \sqrt{-g}d\tau$ and integrated over $\Sigma$ and obtained:

$$(3)\, \int \sqrt{-g'}\, \delta g'^{\rho\sigma} d\tau' = \int \sqrt{-g}\, d\tau \sum \frac{\partial x'_\rho}{\partial x_\mu} \frac{\partial x'_\sigma}{\partial x_\nu} \delta g^{\mu\nu}$$

The integration region $\Sigma$ was infinitely small and so Einstein substituted the factors, $\frac{\partial x'_\rho}{\partial x_\mu} \frac{\partial x'_\sigma}{\partial x_\nu}$ with the constant values that these factors obtain for any one place in the integration space.

He then re-wrote (3) in the following short form. He neglected infinitesimally small terms on the right-hand of (3) and wrote the following equation (integrated over the domain $\Sigma$):

$$(4)\, A^{\mu\nu} = \int \sqrt{-g}\, \delta g^{\mu\nu} d\tau.$$

It is equation (8) the transformation law for contravariant tensors, from the 1914 paper: [33]

$$(3a)\, A'^{\rho\sigma} = \sum \frac{\partial x'_\rho}{\partial x_\mu} \frac{\partial x'_\sigma}{\partial x_\nu} A^{\mu\nu}.$$

Einstein told Levi-Civita that **$A^{\mu\nu}$ was a contravariant tensor and, more precisely one whose components can be chosen independently from one another.**

Einstein then reformulated equation (1). He said that since [(74)] is a "slowly" variable function of the coordinates, then one could just consider it as constant in the right hand side of equation (1). Einstein thus set:

$$\delta I = \sum_{\mu\nu} \left\{ \frac{\mathfrak{G}_{\mu\nu}}{\sqrt{-g}} \int \sqrt{-g}\, \delta g^{\mu\nu} d\tau \right\},$$

and with (4) this led to,

$$(1a)\ \delta I = \sum \left\{ \frac{\mathfrak{G}_{\mu\nu}}{\sqrt{-g}} A^{\mu\nu} \right\}.$$

Einstein concluded that since according to the assumption $\delta I$ is an invariant, and $A^{\mu\nu}$ is a contravariant tensor with independent chosen components, then so is:

$$\frac{\mathfrak{G}_{\mu\nu}}{\sqrt{-g}}$$

a covariant tensor.

Einstein's confidence in this proof was not 100%, because he ended his letter with the following sentence, "In the cheerful hope that you will not find any significant holes in this proof […]".[34]

**The Third letter from March 20, 1915: the constant $g_{\mu\nu}$'s**

Einstein was disappointed. Levi-Civita was still not satisfied. Einstein received Levi-Civita's letter with a counterargument that disproved the tensorial character of [(74)],

If [(74)] is a tensor, then it follows that if it vanishes in one adapted coordinate system, it should vanish in *all* adapted coordinate systems. When the $g_{\mu\nu}$'s are not constant then [(74)] vanishes in the adapted coordinate system. However, if one transforms the field $g_{\mu\nu}$ to the Newtonian case then [(74)] does not vanish. In both cases [(74)] should have vanished.

Einstein did not accept Levi-Civita's criticism, and said,[35]

"I feel this is not correct. It should be kept in mind that, it is generally not possible to change by transformation of the coordinates any given $g_{\mu\nu}$ field into one in which the $g_{\mu\nu}$'s are constant. It will always be impossible to be done, for ex., for parts of a Newtonian field, which generally contain masses, incidentally not even for mass-free regions.
I believe that the proof or disproof of the sentence quoted from your letter is just as difficult as the proof or disproof of the general sentence of the tensor character of […] [(74)]."

## The fourth letter from March 26, 1915: $A^{\mu\nu}$'s can be chosen independently of one another

Levi-Civita was going to show Einstein on March 28 exactly what he considered impossible. Meanwhile, three days later Levi-Civita responded, finally in Italian: "I have just received your letter of March 23$^{rd}$ written in the so-familiar Italian, of which I have been deprived for long time", wrote Einstein.[36]

Einstein using his usual sense of humor mixed with cynicism and honesty wrote Levi-Civita, "You also do it very pretty in your letters: first you flatter me kindly, to prevent me from making a dour face upon reading your new objections".[37]

It appears that Levi-Civita waited for a reply to his previous objections – upon restriction to adapted coordinate systems there exist among them systems whose $g_{\mu\nu}$'s are constant. However, Einstein did not supply an answer. Thus Levi-Civita very likely raised this objection again and this time Einstein said, "once again to the objection", and he started to deal with it.[38]

Already on March 20 Einstein quoted Levi-Civita telling him, "This tensor is not identical to zero for all adapted coordinate systems; this is particularly evident in Newton's case".[39]

Einstein could not accept such a statement because according to his equation, $(42a) \sum_{\nu} \frac{\partial \mathfrak{T}_{\tau}^{\nu}}{\partial x_{\nu}} = \frac{1}{2} \sum_{\mu\tau\nu} g^{\tau\mu} \frac{\partial g_{\mu\nu}}{\partial x_{\sigma}} \mathfrak{T}_{\tau}^{\nu} + \mathfrak{K}_{\sigma}$, the divergence of the energy tensor $\mathfrak{T}_{\tau}^{\nu}$ vanishes for all adapted coordinate systems, and the same applies – according to his field equations $\mathfrak{G}_{\sigma\tau} = \chi \mathfrak{T}_{\sigma\tau}$ to the tensor $\mathfrak{G}_{\mu\nu}$.[40]

Hence Einstein wrote Levi-Civita: "But you did not justify this statement, and I am considering it incorrect as long as you have not provided an example or a general proof for this".[41] By the beginning of April Einstein received the justification in the form of a concrete example (discussed below).[42]

Before that, Levi-Civita very likely told Einstein that he had not yet received an answer to another objection raised by him in his previous letter to Einstein. Levi-Civita could not accept Einstein's equation (4) from his March 17 letter, and his assumption that $A^{\mu\nu}$ was a contravariant tensor whose components can be chosen independently from one another. Therefore he objected to equation (1a) from this letter. On March 23, Levi-Civita corrected (1a) to,

$$(1b) \sum_{\mu\nu} \frac{1}{\sqrt{-g}} \mathfrak{G}_{\mu\nu} A^{\mu\nu} + \varepsilon = \text{invariant,}$$

where, $\varepsilon$ is an infinitesimal quantity of higher order, nevertheless [(74)] was not a tensor.[43]

Einstein responded that he did not quite understand this objection of Levi-Civita, but he was going to prove that [(74)] was nevertheless a tensor when (1b) is invariant.

Einstein then tried to persuade Levi-Civita that for any justified substitutions (substitutions between adapted coordinate systems) and up to relatively infinitely small quantities:

$$(1) \sum \frac{1}{\sqrt{-g}} \mathfrak{G}_{\mu\nu} A^{\mu\nu} = \sum \frac{1}{\sqrt{-g'}} \mathfrak{G}'_{\sigma\tau} A'^{\sigma\tau}.$$

Einstein indeed did not quite understand Levi-Civita's objection, because he again wrote equation (3a) from his March 17 letter: [44]

$$(2)\ A'^{\sigma\tau} = \sum \frac{\partial x'_\sigma}{\partial x_\mu} \frac{\partial x'_\tau}{\partial x_\nu} A^{\mu\nu}.$$

Using (1) and (2) it follows that up to infinitely small quantities:

$$[3] \sum_{\mu\nu} \frac{\mathfrak{G}_{\mu\nu}}{\sqrt{-g}} A^{\mu\nu} = \sum_{\sigma\tau\mu\nu} \frac{\mathfrak{G}'_{\sigma\tau}}{\sqrt{-g'}} \frac{\partial x'_\sigma}{\partial x_\mu} \frac{\partial x'_\tau}{\partial x_\nu} A^{\mu\nu}$$

Applies. And assuming **the independence of the $A^{\mu\nu}$'s**:

$$[4] \frac{\mathfrak{G}_{\mu\nu}}{\sqrt{-g}} = \sum_{\sigma\tau} \frac{\partial x'_\sigma}{\partial x_\mu} \frac{\partial x'_\tau}{\partial x_\nu} \frac{\mathfrak{G}'_{\sigma\tau}}{\sqrt{-g'}}.$$

He thus concluded that $\frac{\mathfrak{G}_{\mu\nu}}{\sqrt{-g}}$ has a tensor character. Einstein thought his derivation was independent of any limit that would replace the $A^{\mu\nu}$ (such as $A^{\mu\nu} + \varepsilon$), equation (1b) of Levi-Civita.[45]

### The fifth letter from March 28, 1915: $\mathfrak{G}_{\mu\nu}$ does not vanish for the Newtonian case

On March 28, 1915 Levi-Civita wrote to Einstein that yesterday (on the 27[th]) he received his "postcard" from the 20[th].[46] This letter of Levi-Civita is the only letter that has survived; on April 2, when Einstein replied to this letter, he enclosed Levi-Civita's letter so that he could refer to it while answering to his proof.[47] Thanks to Einstein enclosing this letter and returning it back to Levi-Civita, the latter's letter was saved.

Levi-Civita wrote to Einstein,

"In your view, my observation [that there are adapted coordinate systems for which tensor $\mathfrak{G}_{\mu\nu}$ does not vanish, when the $g_{\mu\nu}$'s are constant] is not conclusive, because a generic gravitational field cannot be obtained with the help of a coordinate transformation starting from a Euclidean $ds^2$ ($g_{\mu\nu}$ constant).

This is indeed the case. Therefore it was necessary to give a concrete example in which not all $\mathfrak{G}_{\mu\nu}$'s vanish as a result of some admissible transformation from a Euclidean $ds^2$ (contrary to what covariance would require)".[48]

Levi-Civita started from a coordinate system in which $ds^2$ has the form:

$$ds^2 = dx_1^2 + dx_2^2 + dx_3^2 + dx_4^2.$$

He then proceeded to perform an infinitesimal transformation, putting:

(1) $x'_\mu = x_\mu + y_\mu$,

where the $y_\mu$ designate any infinitesimal functions of x.

He put,

(2) $\delta_{\mu\nu} + h_{\mu\nu}$,

while the fundamental tensor $g_{\mu\nu} = \delta_{\mu\nu}$ relative to the new variables x', and,

(3) $h_{\mu\nu} = -\left(\dfrac{\partial y_\mu}{\partial x_\nu} + \dfrac{\partial y_\nu}{\partial x_\mu}\right).$

He used Einstein's 1914 expression for H,[49]

$$H = \frac{1}{4}\sum_{\alpha\beta\tau\rho} g^{\alpha\beta} \frac{\partial g_{\tau\rho}}{\partial x_\alpha} \frac{\partial g^{\tau\rho}}{\partial x_\beta},$$

and obtained for the gravitation tensor,

(4) $\mathfrak{G}_{\mu\nu} = -\dfrac{1}{2}\Delta_2 h_{\mu\nu}$ $\left[\Delta = \sum\limits_{1}^{4} \sigma\dfrac{\partial^2}{\partial x_\sigma^2}\ \text{is the Laplace operator}\right].$

The infinitesimal transformation (1) would be adapted (to a coordinate system) if the y's satisfied Einstein's condition (65a) $B_\mu = 0$, which in the present case became:

(5) $\dfrac{1}{2}\sum\limits_{\nu} \dfrac{\partial}{\partial x_\nu} \Delta_2 h_{\mu\nu} = 0.$

Levi-Civita assumed that in equation (1) $y_\mu = 1/6\, c_\mu x^3{}_\mu$, with constant $c_\mu$. Hence,

$$h_{\mu\mu} = -c_\mu x^2_\mu, \quad \text{And,} \quad \Delta_2 h_{\mu\mu} = -2c_\mu.$$

Inserting this into (4) leads to non-vanishing:

$$\mathfrak{G}_{\mu\mu} = +c_\mu \neq 0.\ ^{50}$$

**The sixth letter from April 2, 1915: modification of covariance proof, only $\delta_1 g^{\mu\nu}$ variations are used**

On April the second Einstein replied. He told Levi-Civita that his letter from March 28 "was extraordinary interesting for me. I had to ponder incessantly for one and a half days until it was clear to me how to reconcile your example with my proof".

As said above, Einstein enclosed Levi-Civita's letter so that he could refer to it while answering to his proof. Einstein first admitted that Levi-Civita's "deduction is entirely correct. $\frac{\mathfrak{G}_{\mu\nu}}{\sqrt{-g}}$ envisaged by you does *not* have a tensor character with the infinitesimal transformation, nevertheless the transformation is performed from a justified coordinate system".[51]

Did Einstein back down from his most important proof in his 1914 review paper? No, he did not do so, because immediately after saying the above, he wrote, "Strangely enough, my proof is *not* refuted by this for the following reason: My proof fails in precisely that special case you have treated".[52]

And so for the time being Einstein continued to adhere to his proof, his Hole Argument and to the restriction to adapted coordinate systems.

Einstein said that if the $\delta g^{\mu\nu}$ are freely chosen, the tensor character of [(74)] follows from the condition [(67)],

(1) $B_\mu = 0$ and $\Delta B_\mu = 0$,

and from (71) (which is invariant) for the infinitesimal substitution under examination. It was enough that the integrals:

$$\int \sqrt{g}\, \delta g^{\mu\nu} d\tau$$

would fulfill this requirement. **He reminded Levi-Civita that he proved earlier that these have a tensor character.** Einstein meant his letter from March 17 where he derived equation **(4)** $A^{\mu\nu} = \int \sqrt{-g}\, \delta g^{\mu\nu} d\tau$, and assumed **the independence of the $A^{\mu\nu}$'s**.[53]

Einstein explained that for a given infinitesimal region, one must be able to choose the integrals $\int \delta g^{\mu\nu} d\tau$ freely for the tensorial character of [(74)] to follow from: (1) $B_\mu = 0$ and $\Delta B_\mu = 0$.

But in the special case examined by Levi-Civita –from his letter of March 28 – the integrals $\int \delta g^{\mu\nu} d\tau$ *could not* be freely chosen, and thus the tensorial character of $\frac{G_{\mu\nu}}{\sqrt{-g}}$ did not follow from (1).

Einstein next went on to show that for Levi-Civita's special case, the $\int \delta g^{\mu\nu} d\tau$ could not be freely chosen because they all vanished.

In order to demonstrate his claim Einstein wrote equation (69) from section §14 of his paper. [54] The $\delta_1 g^{\mu\nu}$ must satisfy the condition: $\delta_1 B_\mu = 0$.

In the special Newtonian case considered by Levi-Civita this condition is reduced to the following form:

$$\sum_\nu \frac{\partial}{\partial x_\nu}(\Delta \delta_1 g^{\mu\nu}) = 0.$$

And Einstein told Levi-Civita to compare this equation to his own equation (5):

$$(5) \frac{1}{2} \sum_\nu \frac{\partial}{\partial x_\nu} \Delta_2 h_{\mu\nu} = 0.$$

(and that was the reason for why Einstein enclosed Levi-Civita's letter from March 28 to this letter as reference).

Because of the boundary conditions,

$$\sum \frac{\partial \delta_1 g^{\mu\nu}}{\partial x_\nu} = 0.$$

Einstein then multiplied this equation by $x_\sigma$, and integrated over the whole region. Then after partial integration of each index combination one obtained,

$$(I) \int \delta_1 g^{\mu\sigma} d\tau = 0.$$

This equation is a consequence of the special form of $\delta_1 B_\mu = 0$ in the case considered by Levi-Civita in his March 28 letter.

Einstein said that it follows from equation (63) from his 1914 paper, [55] and from the definition of the $\delta_2 g^{\mu\nu}$'s for an infinitely small region that: [56]

$$(II) \int \delta_2 g^{\mu\nu} d\tau = 0.$$

Thus from (I) and (II) it follows for each index combination:

$$\int \delta g^{\mu\nu} d\tau = 0.$$

By this Einstein showed that for Levi-Civita's special case, $\int \delta g^{\mu\nu} d\tau$ did indeed vanish.

Einstein concluded his letter by saying, [57]

"It follows that in the specialization you have introduced, it is inherent to it that the equations (1) of this letter are *not* sufficient conditions for a tensor character of $\frac{\mathfrak{G}_{\mu\nu}}{\sqrt{-g}}$ under an infinitesimal transformation.

Generally, however, equation $\delta_1 B_\mu = 0$ cannot be reduced to a first-order equation for the $\delta_1 g^{\mu\nu}$'s. Then my proof correctly holds for all finite transformations".

This supplied a somewhat formal basis to Einstein's March 20 reasoning: "It should be kept in mind that, it is generally not possible to change by transformation of the coordinates any given $g_{\mu\nu}$ field into one in which the $g_{\mu\nu}$'s are constant. It will always be impossible to be done, for ex., for parts of a Newtonian field, which generally contain masses, incidentally not even for mass-free regions".[58]

However, there was a little problem in the above derivation. Einstein inspected that his observation that his proof holds for all finite transformations, suggested a modification of his covariance proof in which only $\delta_1 g^{\mu\nu}$ variations would be used since the $\delta_2 g^{\mu\nu}$ do not actually contribute anything to $\int \sqrt{g}\delta g^{\mu\nu} d\tau$.[59]

**The seventh letter from April 8, 1915: the quasi constants γ^μν**

And indeed again Levi-Civita was not satisfied with Einstein's reply of $\int \delta g^{\mu\nu} d\tau = 0$ from the previous letter. On the very same day, April the second, he sent Einstein a postcard with the same objection; and Einstein tried again to disprove Levi-Civita's objection on April 8, 1915, this time by using a different strategy: the quasi constant $g_{\mu\nu}$'s. Einstein did it "more in order gradually to learn the main focus of your reservations".[60]

Einstein explained to Levi-Civita that with respect to his special case from March 28: "My proof of the invariant nature of δJ fails with such infinitesimal transformations, in which the $g_{\mu\nu}$'s of the original system are constant, because then the quantities $A^{\mu\nu}$ cannot be chosen freely, but vanish altogether".[61] Einstein was stubbornly clinging to what has remained from his proof, and he now thought that it did not fail generally, but only in certain special cases; and Levi-Civita's case falls exactly on such a special case in which the $g_{\mu\nu}$'s of the system are constant. Einstein thought that at this stage, "it proves nothing about the validity of the theorem in general".[62]

On April 8 Einstein came up with an ingenious new idea: he showed that although his proof failed with constant $g^{\mu\nu}$, it did not fail with quasi constant $g^{\mu\nu}$!

And the tensor $\frac{\mathfrak{G}_{\mu\nu}}{\sqrt{-g}}$ was covariant; hence $\mathfrak{G}_{\mu\nu}$ has a tensor character and it could be used as a gravitation tensor.

Consider a finite zone $\Sigma$. Einstein selected in this zone an inner region $\sigma$, within which he determined $\delta g^{\mu\nu}$ as equal to constant $\gamma^{\mu\nu}$, whose value remains unchanged; but a boundary zone brings the gradual transition to zero. Then he said that there was a break which was equal to: $(\Sigma - \sigma)/\Sigma$, and could be made arbitrarily small. In the proof Einstein proceeded to $\Sigma = 0$ such that he completed the transition at constant $\gamma^{\mu\nu}$ and constant $(\Sigma - \sigma)/\Sigma$.

Einstein took his equation **(4)** $A^{\mu\nu} = \int \sqrt{-g}\,\delta g^{\mu\nu}\,d\tau$ from his letter to Levi-Civita from March 17 (an equation to which Levi-Civita objected) and obtained for $\delta g^{\mu\nu}$ equal to constant $\gamma^{\mu\nu}$:[63]

$$A^{\mu\nu} = \int \sqrt{-g}\,\delta g^{\mu\nu}\,d\tau = \sqrt{-g}\,\gamma^{\mu\nu}\,\Sigma.$$

Therefore,

$$(1)\ A^{\mu\nu} = \int \delta g^{\mu\nu}\,d\tau = \gamma^{\mu\nu}\,\Sigma.$$

The smaller $(\Sigma - \sigma)/\Sigma$ was chosen, the more accurate the above equation was satisfied.

Einstein returned to equations from his letter from March 26. Recall that in this letter Einstein said that for infinitely small quantities,

$$[3]\ \sum_{\mu\nu} \frac{\mathfrak{G}_{\mu\nu}}{\sqrt{-g}} A^{\mu\nu} = \sum_{\sigma\tau\mu\nu} \frac{\mathfrak{G}'_{\sigma\tau}}{\sqrt{-g'}} \frac{\partial x'_\sigma}{\partial x_\mu} \frac{\partial x'_\tau}{\partial x_\nu} A^{\mu\nu}$$

applies. Assuming **the independence of the $A^{\mu\nu}$'s,** Einstein concluded that equation

$$[4]\ \frac{\mathfrak{G}_{\mu\nu}}{\sqrt{-g}} = \sum_{\sigma\tau} \frac{\partial x'_\sigma}{\partial x_\mu} \frac{\partial x'_\tau}{\partial x_\nu} \frac{\mathfrak{G}'_{\sigma\tau}}{\sqrt{-g'}}$$

applies.[64]

On April 8, Einstein still did not realize that [3] did not lead to [4]; he only thought that there was one specific case – Levi-Civita's case from March 28 – which disturbs his proof. And so Einstein wrote [3] again. However, he now divided equation [3] by the region $\int d\tau = \Sigma$ and then selected the $\delta g^{\mu\nu}$ according to his new determination of constants $\gamma^{\mu\nu}$. He then further claimed that with the aid of the expression (1) for the $A^{\mu\nu}$, and the smaller $(\Sigma - \sigma)/\Sigma$ was, he could arrive with "greater accuracy" to:

$$[4a] \sum \frac{\mathfrak{G}_{\mu\nu}}{\sqrt{-g}} \gamma^{\mu\nu} = \sum \frac{\mathfrak{G}'_{\sigma\tau}}{\sqrt{-g'}} \frac{\partial x'_\sigma}{\partial x_\mu} \frac{\partial x'_\tau}{\partial x_\nu} \gamma^{\mu\nu}.$$

Since the $\gamma^{\mu\nu}$ could be chosen arbitrarily, then one arrives at [4].[65]

On March 26 Einstein had already concluded from equation [4] that $\frac{\mathfrak{G}_{\mu\nu}}{\sqrt{-g}}$ has a tensor character.[66]

Subsequently, Einstein told Levi-Civita, "You have received my letter refuting your example. I shall repeat myself".[67] And he explained to him the general reasoning brought above before this proof. Although Einstein admitted that his proof of the invariance of δJ fails, he found a subtle way to save his proof in the face of Levi-Civita's objections.

And Einstein was so confident that he closed his letter by saying, "I must even admit that, through the deeper considerations to which your interesting letters have led me, I have become even more firmly convinced that the proof of the tensor character of $\frac{\mathfrak{G}_{\mu\nu}}{\sqrt{-g}}$ is correct in principle".[68]

**The eighth letter from April 11, 1915: H is invariant to arbitrary substitutions and thus to linear ones**

On April 11 Einstein sent Levi-Civita a postcard, in which he wrote a few sentences: "You get a nice special case for the claim that $\frac{\mathfrak{G}_{\mu\nu}}{\sqrt{-g}}$ is a tensor when you set H = const. The condition $B_\mu = 0$ is then satisfied identically, and H is invariant under arbitrary, and thus also linear substs. [Substitutions]. $\frac{\mathfrak{G}_{\mu\nu}}{\sqrt{-g}}$ must, according to the theorem in this case, have a tensor character under arbitrary substitutions. Indeed results,

$$\frac{\mathfrak{G}_{\mu\nu}}{\sqrt{-g}} = \text{const} \cdot g_{\mu\nu}$$

therefore truly a covariant tensor".[69]

Cattani and De Maria said that this choice of H = constant is incompatible with Einstein's previous choice from his 1914 paper – which according to Einstein – led uniquely to his old "Entwurf" field equations in this paper.[70] Einstein was thus confused; and all he cared now was that $\frac{\mathfrak{G}_{\mu\nu}}{\sqrt{-g}}$ would be a "wirklich" covariant tensor.

**The ninth letter from April 14, 1915: Einstein would acknowledge Levi-Civita's efforts when repeating the $\mathfrak{G}_{\mu\nu}$ −proof**

Levi-Civita replied with a postcard and was still not satisfied. On April 14 Einstein wrote Levi-Civita again. Einstein answered him, "When the opportunity arises to repeat the $\mathfrak{G}_{\mu\nu}$ −proof, I shall gladly include the improvements I have learned from our memorable correspondence".[71]

At this stage Einstein thought he would repeat his proof from his 1914 review paper that his gravitational tensor is a covariant tensor. On April 14, 1915, Einstein was therefore quite sure that he would hold the "Entwurf" equations, and would need this proof to future papers on the subject.

**The tenth letter from April 20, 1915: the quasi constants again**

On April 15[th], a day afterwards, Einstein received another letter from Levi-Civita. Levi-Civita did not agree with Einstein's April 8 choice of the $\delta g^{\mu\nu}$ as equal to constant $\gamma^{\mu\nu}$.[72] To this Einstein replied five days later, "but I acknowledge you that you have put your finger on the weakest point of the proof, namely, on the independence of the $A^{\mu\nu}$'s".[73] Einstein was still stubbornly clinging to his $A^{\mu\nu}$'s and he said, "here the proof lacks precision; the theorem on page 1072 [of his 1914 paper] 'therefore there will be equivalent to an arbitrary variation of the $\delta g^{\mu\nu}$'s' dispenses with rigorous justification, even in the special case of a constant $g^{\mu\nu}$ it is incorrect. But I am still firm confident that it is generally true, just because the number of freely chosen variables that determine the 10 $\delta g^{\mu\nu}$'s is 10, and because both variations $\delta_1$ and $\delta_2$ are of fundamentally different kinds, in that a $\delta_2$ variation generally is not a $\delta_1$ variation".[74]

When examining Levi-Civita's special case, the $g^{\mu\nu}$ were completely constant and the $\delta g^{\mu\nu}$ were not freely chosen. Einstein realized that only $\delta_1 g^{\mu\nu}$ variations should be used since the $\delta_2 g^{\mu\nu}$ do not actually contribute anything to $\int \sqrt{g} \delta g^{\mu\nu} d\tau$.[75]

In his letter to Einstein from April 15, Levi-Civita told Einstein "right from the beginning" that for small regions Σ:[76]

$$A^{\mu\nu} = \int \delta g^{\mu\nu} d\tau = \int \delta_1 g^{\mu\nu} d\tau,$$

and the terms $\int \delta_2 g^{\mu\nu} d\tau$ vanish. And Levi-Civita deduced that $\int \delta_2 g^{\mu\nu} d\tau$ vanish. Therefore, if only $\delta_1 g^{\mu\nu}$ variations should be used (on April 2 Einstein indeed said this), then unlike Einstein's claim from above, the number of freely chosen variables that determine the 10 $\delta g^{\mu\nu}$'s is not 10. The choice of the $\delta g^{\mu\nu}$ as equal to constant $\gamma^{\mu\nu}$ did not solve this problem.

Einstein's reaction was denial, "Schon dies bestreite ich".[77]

**The eleventh letter from April 21, 1915: the comeback of the constant H**

A day afterwards Einstein sent Levi-Civita another letter answering a letter that had probably arrived a day before. Einstein returned to another dubious argument from his April 11, 1915 letter. Recall he there wrote that one gets a special case for the statement that $\frac{\mathfrak{G}_{\mu\nu}}{\sqrt{-g}}$ is a tensor when one sets H = const. The condition $B_\mu = 0$ is then satisfied identically, and H is invariant for arbitrary substitutions. Einstein said that $\frac{\mathfrak{G}_{\mu\nu}}{\sqrt{-g}} = \text{const} \cdot g_{\mu\nu}$, is a covariant tensor.[78]

Levi-Civita thought that Einstein sent him this as an answer to his arguments against Einstein persistence on the independence of the $A^{\mu\nu}$'s. Einstein therefore wrote Levi-Civita, "I gave you the example H = const. not related to refute your new objection (conc. the independence of the $A^{\mu\nu}$'s), but to refute the original objection [that $\frac{\mathfrak{G}_{\mu\nu}}{\sqrt{-g}}$ was not a tensor]".[79]

Einstein realized that Levi-Civita still objected to his claims and he said "As to your postcard, I see that you attach great importance to your *new* objections, which culminate in the statement of the vanishing of the $A^{\mu\nu}$'s, by again finding that the theorem does not hold true. But I hope that the letter I sent yesterday[80] will convince you",[81] but it did not convince Levi Civita who was stubborn in his objection.

**The twelfth and final letter from May 5, 1915: $A^{\mu\nu}$'s cannot be chosen arbitrarily**

The final surviving letter of Einstein to Levi-Civita is from May 5, 1915. Einstein summarized, "I also believe that we have exhausted our subject so far, inasmuch our present state of the same knowledge allows. My proof is incomplete to the extent that it is not proven that the $A^{\mu\nu}$'s can be chosen arbitrarily".[82]

Einstein was disappointed and he then desperately wrote Levi-Civita that his formulation for a possible gravitation tensor:

$$\mathfrak{G}_{\mu\nu} = \frac{1}{2} g_{\mu\nu} \Delta\varphi - \frac{\partial\varphi}{\partial x_\mu} \frac{\partial\varphi}{\partial x_\nu}$$

cannot solve the problem of the relativity of motion. This was so, because if φ is constant for one coordinate system, then it is constant for all the others as well. And so no moving coordinate system would have a gravitational field – which is impossible.[83]

## 3. Conclusions

Cattani and De Marria wrote, "In particular, the main bug that Einstein discovered with the help of Levi-Civita (which he called the 'sorest spot' of his proof) regards the question of the independence of the $A^{\mu\nu}$. […] Einstein was obliged to admit that his demonstration was 'incomplete, in the sense that the possibility of an arbitrary choice of $A^{\mu\nu}$ remains unproved'. This conclusion, as we have shown, implies the impossibility of proving the tensorial character of $\frac{\mathfrak{G}_{\mu\nu}}{\sqrt{-g}}$ within a limited-covariance theory".[84]

Recall that on April 21, Einstein wrote to Levi-Civita, "I gave you the example H = const. not related to refute your new objection (conc. the independence of the $A^{\mu\nu}$'s), but to refute the original objection [that $\frac{\mathfrak{G}_{\mu\nu}}{\sqrt{-g}}$ was not a tensor]".[85]

Immediately afterwards, in the final surviving letter from May 5, Einstein accepted Levi-Civita's *new objection* when he said his proof was incomplete to the extent that it was not proven that the $A^{\mu\nu}$'s could be chosen arbitrarily[86]. However, Einstein still did not fully understand why $\frac{\mathfrak{G}_{\mu\nu}}{\sqrt{-g}}$ was not a tensor. He received an answer to this question from another brilliant mathematician, David Hilbert from Göttingen.

In autumn 1915 Hilbert found a flaw in Einstein's proof from section §14 of his 1914 review paper (on November 7, 1915, Einstein wrote to Hilbert, "you found a hair in my soup, which spoiled it entirely for you")[87].

In October 1915, Einstein, "realized, namely, that my existing field equations of gravitation were entirely untenable!"[88] And in November 1915 he arrived back at generally covariant field equations, from which he parted with a heavy heart three years earlier.[89] But only in 1916 Einstein understood the error which Hilbert found in his proof.

On March 30, 1916, Einstein sent to Hilbert a letter in which he explained the mistake,[90] "The error you found in my paper of 1914 has now become completely clear to me".[91]

After writing equation [(61)] from his 1914 paper, Einstein wrote the following equation: $\delta g_\sigma^{\mu\nu} = \frac{\partial}{\partial x_\sigma}(\delta g^{\mu\nu})$. Taking this into consideration Einstein wrote equation (71) and then (72) and (73); the latter equation was his gravitational tensor $\mathfrak{G}_{\mu\nu}$. Hilbert told Einstein that the above equation is not valid.[92] Einstein understood that if the above equation is not valid, then from the mathematical point of view, equation (71), equation (72) and Einstein's gravitational tensor $\mathfrak{G}_{\mu\nu}$ are all not valid. *Hence only in spring 1916, after he had already abandoned his 1914 proof and gravitational tensor $\frac{\mathfrak{G}_{\mu\nu}}{\sqrt{-g}}$, did Einstein understand his mistake and why $\frac{\mathfrak{G}_{\mu\nu}}{\sqrt{-g}}$ was not a tensor.*

I wish to thank Prof. John Stachel from the Center for Einstein Studies in Boston University for sitting with me for many hours discussing special and general relativity and their history. Almost every day, John came with notes on my draft manuscripts, directed me to books in his Einstein collection, and gave me copies of his papers on Einstein, which I read with great interest. I also wish to thank Prof. Alisa Bokulich, Director of the Boston University Center for History and Philosophy of Science, for her kind assistance while I was a guest of the Center. Finally I would like to thank her Assistant, Dimitri Constant, without whose advice and help I would not have been able to get along so well at BU and in Boston in general!

## Endnotes

[1] Martínez, Alberto. A, "The Myriad Pieces of Einstein's Remains", Annals of Science 68, 2011, pp. 267-280; p. 277.

[2] *The Collected Papers of Albert Einstein. Vol. 8: The Berlin Years:Correspondence, 1914–1918* (*CPAE*, Vol. 8), Schulmann, Robert, Kox, A.J., Janssen, Michel, Illy, Jószef (eds.), Princeton: Princeton University Press, 2002.

[3] Einstein to Levi-Civita, March 5, 1915, *CPAE*, Vol. 8, Doc. 60; Einstein to Levi-Civita, March 17, 1915, *CPAE*, Vol. 8, Doc. 62; Einstein to Levi-Civita, March 20, 1915, *CPAE*, Vol. 8, Doc. 64; Einstein to Levi-Civita, March 26, 1915, *CPAE*, Vol. 8, Doc. 66; Levi-Civita to Einstein, March 28, 1915, *CPAE*, Vol. 8, Doc. 67; Einstein to Levi-Civita, April 2, 1915, *CPAE*, Vol. 8, Doc. 69; Einstein to Levi-Civita, April 8, 1915, *CPAE*, Vol. 8, Doc. 71; Einstein to Levi-Civita, April 11, 1915, *CPAE*, Vol. 8, Doc. 74; Einstein to Levi-Civita, April 14, 1915, *CPAE*, Vol. 8, Doc. 75; Einstein to Levi-Civita, April 20, 1915, *CPAE*, Vol. 8, Doc. 77; Einstein to Levi-Civita, April 21, 1915, *CPAE*, Vol. 8, Doc. 78; Einstein to Levi-Civita, May 5, 1915, *CPAE*, Vol. 8, Doc. 80.

[4] Cattani, Carlo and De Maria Michelangelo, "The 1915 Epistolary Controversy between Einstein and Tullio Levi-Civita", in Howard, Don and Stachel, John (eds.), *Einstein and the History of General Relativity: Einstein Studies, Volume 1*, 1989, New York: Birkhauser, pp. 175-200; p. 176.

[5] Einstein, Albert, "Die formale Grundlage der allgemeinen Relativitätstheorie", *Königlich Preußische Akademie der Wissenschaften* (Berlin). *Sitzungsberichte*, 1914, pp. 1030-1085.

[6] Following Ricci and Levi-Civita, Einstein explained that he represented contravariant components of a tensor by raised indices and covariant components by lowered indices.

[7] Einstein, 1914, p. 1069.

[8] Einstein, 1914, pp. 1041-1042. The line element:

$$ds^2 = \sum_{\mu\nu} g_{\mu\nu} dx_\mu dx_\nu = \sum_\sigma dX_\sigma^2$$

The ten quantities $g_{\mu\nu}$ determine the gravitational field and they are functions of $x_\nu$, and special relativity still holds in the infinitesimally small. It follows that the proper time:

$$d\tau_0^* = \int dX_1\, dX_2\, dX_3\, dX_4$$

over a volume element is an invariant.
Einstein wrote the following relation for $g^{\mu\nu}$:

$$(17)\ \sqrt{g} d\tau = d\tau_0^*.$$

And he rewrote it in the following form:

$$(17a)\ \sqrt{-g} d\tau = d\tau_0.$$

Equation (17) could be interpreted as the relation between the proper time, $d\tau_0^{\bullet}$, which is measured by a clock at rest with the observer along his worldline, and the coordinate time $\sqrt{g}d\tau$. Equation (17a) is interpreted in the following way: the $dX_\sigma$ correspond to the customary coordinates in special relativity. Three of these coordinates are real-valued, and one is imaginary, $dX_4$. Consequently, $d\tau_o$ is imaginary.

[9] By taking the following relations into account:
$\Delta g^{\mu\nu} = g^{\mu\nu'} - g^{\mu\nu}$,
$\Delta x_\mu = x'_\mu - x_\mu$.

[10] Einstein, 1914, p. 1069.

[11] Einstein, 1914, p. 1070.

[12] Einstein, 1914, p. 1070.

[13] Einstein, 1914, p. 1071.

[14] Einstein, 1914, p. 1071.

[15] Einstein, 1914, pp. 1071-1072.

[16] Einstein, 1914, p. 1073.

[17] Einstein, 1914, p. 1071.

[18] Einstein, 1914, p. 1073.

[19] Einstein, 1914, pp. 1073-1074.

[20] Cattani and De Maria, 1989, p. 176.

[21] Einstein to Levi-Civita, April 2, 1915, *CPAE,* Vol. 5, Doc. 69.

[22] Einstein to Levi-Civita, March 5, 1915, *CPAE*, Vol. 8, Doc. 60.

[23] Einstein to Levi-Civita, March 5, 1915, *CPAE*, Vol. 8, Doc. 60.

[24] Einstein to Levi-Civita, March 5, 1915, *CPAE*, Vol. 8, Doc. 60.

[25] Einstein to Levi-Civita, March 5, 1915, *CPAE*, Vol. 8, Doc. 60.

[26] Einstein to Levi-Civita, March 5, 1915, *CPAE*, Vol. 8, Doc. 60.

[27] Einstein, 1914, p. 1073.

[28] Einstein to Levi-Civita, March 5, 1915, *CPAE*, Vol. 8, Doc. 60.

[29] Einstein to Levi-Civita, March 17, 1915, *CPAE*, Vol. 8, Doc. 62.

[30] Einstein to Levi-Civita, March 17, 1915, *CPAE*, Vol. 8, Doc. 62.

[31] Einstein to Levi-Civita, March 17, 1915, *CPAE*, Vol. 8, Doc. 62.

[32] For Einstein's dispute with De Sitter see: Janssen, Michel, "The Einstein-De Sitter Debate and its Aftermath", lecture, pp. 1-8, based on "The Einstein-De Sitter-Weyl-Klein Debate" in *CPAE*, Vol 8, 1998, pp. 351-357.

[33] Einstein, 1914, p. 1037.

[34] Einstein to Levi-Civita, March 17, 1915, *CPAE*, Vol. 8, Doc. 62.

[35] Einstein to Levi-Civita, March 20, 1915, *CPAE*, Vol. 8, Doc. 64.

[36] Einstein to Levi-Civita, March 26, 1915, *CPAE*, Vol. 8, Doc. 66.

[37] Einstein to Levi-Civita, March 26, 1915, *CPAE*, Vol. 8, Doc. 66.

[38] Einstein to Levi-Civita, March 26, 1915, *CPAE*, Vol. 8, Doc. 66.

[39] Einstein to Levi-Civita, March 20, 1915, *CPAE*, Vol. 8, Doc. 64.

[40] Einstein, 1914, p. 1074.

[41] Einstein to Levi-Civita, March 26, 1915, *CPAE*, Vol. 8, Doc. 66.

[42] Levi-Civita to Einstein, March 28, 1915, *CPAE*, Vol. 8, Doc. 67.

[43] Einstein to Levi-Civita, March 26, 1915, *CPAE*, Vol. 8, Doc. 66.

[44] Einstein to Levi-Civita, March 17, 1915, *CPAE*, Vol. 8, Doc. 62.

[45] Einstein to Levi-Civita, March 26, 1915, *CPAE*, Vol. 8, Doc. 66.

[46] Einstein to Levi-Civita, March 20, 1915, *CPAE*, Vol. 8, Doc. 64.

[47] Einstein to Levi-Civita, April 2, 1915, *CPAE*, Vol. 8, Doc. 69.

[48] Levi-Civita to Einstein, March 28, 1915, *CPAE*, Vol. 8, Doc. 67.

[49] Einstein, 1914, pp. 1074-1076.

[50] Levi-Civita to Einstein, March 28, 1915, *CPAE*, Vol. 8, Doc. 67.

[51] Einstein to Levi-Civita, April 2, 1915, *CPAE*, Vol. 8, Doc. 69.

[52] Einstein to Levi-Civita, April 2, 1915, *CPAE*, Vol. 8, Doc. 69.

[53] Einstein to Levi-Civita, March 17, 1915, *CPAE*, Vol. 8, Doc. 62.

[54] Einstein, 1914, p. 1071.

[55] Einstein, 1914b, p. 1069.

[56] Explained in Cattani and De Maria, 1989, in Howard and Stachel (1989), p. 198, note 49.

[57] Einstein to Levi-Civita, April 2, 1915, *CPAE*, Vol. 8, Doc. 69.

[58] Einstein to Levi-Civita, March 20, 1915, *CPAE*, Vol. 8, Doc. 64.

[59] Einstein to Levi-Civita, April 2, 1915, *CPAE*, Vol. 8, Doc. 69.

[60] Einstein to Levi-Civita, April 8, 1915, *CPAE*, Vol. 8, Doc. 71.

[61] Einstein was adhering to the same reasoning he had presented in his letter from March 17 that $A^{\mu\nu}$ was a contravariant tensor and one whose components can be chosen independently of one another; and to his reasoning from his previous letter from April 2. Einstein to Levi-Civita, March 17, 1915, *CPAE*, Vol. 8, Doc. 62.

[62] Einstein to Levi-Civita, April 8, 1915, *CPAE*, Vol. 8, Doc. 71.

[63] Einstein to Levi-Civita, March 17, 1915, *CPAE*, Vol. 8, Doc. 62.